\documentclass[12pt]{article}
\usepackage{epsfig,amsfonts,amssymb}
\usepackage{hyperref}
\usepackage{comment}
\usepackage{cite}
\input epsf.sty
\usepackage{cite}

\newcommand{\e}[1]{\text{e}^{#1}}

\usepackage{slashed,amssymb,amsfonts,amsmath}
\usepackage{comment}
\usepackage{cancel}
\usepackage{bbm}
\usepackage{array}
\usepackage{color}
\usepackage{graphicx}
\usepackage{mathrsfs}
\usepackage{hyperref}
\usepackage{tikz}
\usepackage{tkz-euclide}
\usetikzlibrary{arrows,calc,positioning,fit,petri,arrows.meta}
\usetikzlibrary{automata,decorations.markings,decorations.pathmorphing}
\usepackage{pgfplots}
\usepackage{tikz-3dplot}
\hypersetup{colorlinks=true}
\hypersetup{linkcolor=blue}
\hypersetup{citecolor=blue}
\hypersetup{urlcolor=blue}
\numberwithin{equation}{section}

\def\e{{\epsilon}}
\def\ve{{\varepsilon}}

\def\g{{\gamma}}

\def\a{{\alpha}}
\def\b{{\beta}}

\def\bz{{\bar z}}

\def\bw{{\bar w}}

\def\da{{\dot \alpha}}
\def\db{{\dot \beta}}

\def\CD{{\mathcal D}}

\def\CI{{\mathcal I}}
\def\ci{{\mathcal I}}
\def\CJ{{\mathcal J}}

\def\CL{{\mathcal L}}

\def\CO{{\mathcal O}}

\def\CQ{{\mathcal Q}}

\def\CS{{\mathcal S}}

\def\CV{{\mathcal V}}

\def\SF{{\mathscr F}}

\def\SS{{\mathscr S}}

\def\p{\partial}
\newcommand{\avg}[1]{\langle \, #1\, \rangle}
\newcommand{\bra}[1]{\langle \,#1 \, |}
\newcommand{\ket}[1]{ | \, #1 \,\rangle}

\def\mrr{{\mathbb R}}

\newcommand{\bd}[1]{\begin{fmffile}{#1}\begin{fmfgraph*}}
\newcommand{\ed}{\end{fmfgraph*}\end{fmffile}}

\def\eg{{e.g.}\ }

\def\0{{(0)}}
\def\1{{(1)}}
\def\2{{(2)}}
\def\3{{(3)}}
\def\+{{(+)}}
\def\-{{(-)}}

\newcommand{\sbra}[1]{  [\, #1 \, | } 
\newcommand{\sket}[1]{  | \, #1  \,] }

\def\<{\langle}
\def\>{\rangle}
\newcommand{\savg}[1]{ [\, #1 \, ]}

\begin{document}

\baselineskip 24pt

\begin{center}
{\Large \bf Asymptotic Symmetries and Subleading Soft Photon Theorem in Effective Field Theories}

\end{center}

\vskip .6cm
\medskip

\vspace*{4.0ex}

\baselineskip=18pt

\centerline{ \rm Alok Laddha$^{a}$ and Prahar Mitra$^{b}$}

\vspace*{4.0ex}

\centerline{ \it ~$^a$Chennai Mathematical Institute, Siruseri, Chennai, India}

\centerline{ \it ~$^b$Institute for Advanced Study, Princeton, NJ, USA }

\vspace*{1.0ex}
\centerline{\small E-mail: aladdha@cmi.ac.in, prahar21@ias.edu}

\vspace*{5.0ex}

\centerline{\bf Abstract} \bigskip

In \cite{Lysov:2014csa,Campiglia:2016hvg} it was shown that the subleading soft photon theorem in tree level amplitudes in massless QED is equivalent to a new class of symmetries of the theory parameterized by a vector field on the celestial sphere. In this paper, we extend these results to the subleading soft photon theorem in any Effective Field Theory containing photons and an arbitrary spectrum of massless particles.  We show that the charges associated to the above class of symmetries are sensitive to certain three point functions of the theory and are corrected by irrelevant operators of specific dimensions. Our analysis shows that the subleading soft photon theorem in any tree level scattering amplitude is a statement about asymptotic symmetries of the $\CS$-matrix.

\bigskip

\vfill \eject

\baselineskip 18pt

\tableofcontents

\section{Introduction}

Soft theorems in gauge theories and gravity have been analyzed in remarkable detail in recent years in light of their close connection with infinite dimensional asymptotic symmetries. In the case of gravity, recent works of Sen \cite{Sen:2017xjn,Sen:2017nim} show that in any UV finite theory of quantum gravity in dimensions greater than four, the leading and subleading soft graviton theorems are universal and completely independent of the matter spectrum of the theory and quantum corrections. The issue is subtler in four dimensions due to the presence of infrared divergences \cite{Bern:2014oka} where the soft factors are IR divergent and need to be regularized. Here the soft theorems are best understood for tree-level amplitudes (where the results of Sen continue to hold). However, such tree-level soft theorems contain their own surprises.
 
 In a recent work \cite{Elvang:2016qvq}, Elvang et al. considered higher derivative corrections to tree-level soft gluon and graviton theorems and showed that the leading soft photon, gluon and graviton theorems \cite{Weinberg:1965nx,Berends:1988zn} and the subleading soft graviton theorem \cite{Cachazo:2014fwa} do not receive higher derivative corrections and are truly universal. On the other hand, the subleading soft photon and gluon theorems \cite{Low:1954kd,GellMann:1954kc,Low:1958sn,Burnett:1967km,Broedel:2014fsa,Bern:2014vva,Bern:2014oka} as well as the sub-subleading soft graviton theorem \cite{Cachazo:2014fwa,Zlotnikov:2014sva} are ``quasi-universal" in that they are corrected by a small class of higher derivative operators (also classified in \cite{Elvang:2016qvq,Laddha:2017ygw}), but the correction has the same kinematic structure that is independent of the operator.

The results of \cite{Elvang:2016qvq} are important in light of the recently understood connection between soft theorems and asymptotic symmetries in Minkowski spacetimes \cite{Strominger:2013lka,Strominger:2013jfa}. The leading soft photon, gluon and graviton theorems have been shown to imply a two-dimensional Kac-Moody symmetry on the celestial sphere \cite{Strominger:2013lka,Strominger:2013jfa,He:2014cra,He:2014laa,Kapec:2014zla,Mohd:2014oja,Kapec:2015ena,He:2015zea,Campiglia:2015qka,Cardona:2015woa,Kapec:2015vwa,Campiglia:2015kxa} whereas the subleading soft graviton theorem implies an infinite-dimensional extension of the BMS symmetry of scattering amplitudes \cite{Kapec:2014opa,Campiglia:2014yka,Kapec:2016jld,Sen:2017nim}. That these soft theorems are uncorrected by higher derivative operators have important implications on universality of the symmetries and possible holographic interpretations of four-dimensional scattering amplitudes in terms of a two-dimensional theory \cite{Cheung:2016iub}. 

On the other hand, the subleading soft photon theorem and the sub-subleading soft graviton theorem do not have the same interpretation in terms of \emph{universal} symmetries (such as gauge transformations and diffeomorphisms) as their leading counterparts and it is  not surprising that they are corrected by higher derivative operators. These soft theorems have also been interpreted as various asymptotic symmetries \cite{Lysov:2014csa,Campiglia:2016hvg,Campiglia:2016efb,Campiglia:2016jdj,Conde:2016rom,Conde:2016csj,Laddha:2017ygw}. 

In particular, in \cite{Campiglia:2016hvg}, it is shown that in the absence of higher derivative interactions, the subleading soft photon theorem is the Ward identity of a certain ``divergent" large $U(1)$ gauge symmetry on the null boundary of Minkowski space, $\CI$. The higher derivative corrections to the subleading soft theorem then present a puzzle in this regard. If it is indeed true that the subleading soft photon theorem is equivalent to a Ward identity, it must then be that the associated charges also receive higher derivative corrections in precisely the right way. That this is the case is a priori not obvious. In general, higher derivative and irrelevant operators fall-off faster near $\ci$ than relevant or marginal operators and therefore do not alter the radiative symplectic structure. How then, could such terms possibly affect the form of the asymptotic charges? In this paper, we resolve this puzzle and show that given the nature of the ``divergent" large gauge transformations that give rise to the universal part of the subleading soft photon theorem, higher derivative operators do induce a correction to the radiative charges. The correction is exactly what is needed to reproduce the soft photon theorem as a Ward identity.

This paper is organized as follows. In \S\ref{sec:elvangreview}, we review the results of Elvang et al. In \S\ref{sec:clreview}, we present our conventions, construct the symplectic form and large gauge charges on $\ci$, and discuss mode expansions for fields on $\ci$. In \S\ref{sec:wardid}, we derive the Ward identity and show that it is implies to the subleading soft photon theorem. In \S\ref{sec:invproof}, we show that the soft theorem implies the Ward identity and therefore prove the complete equivalence of soft theorem and Ward identity.

\section{Higher Derivative Corrections to the Soft Theorems}\label{sec:elvangreview}

In this paper, we study the corrections to the subleading soft theorem in the presence of higher derivative terms. The actions we consider take the general form
\begin{equation}
\begin{split}\label{genaction}
\CL = - \frac{1}{4e^2} F_{\mu\nu} F^{\mu\nu}+ \CL^{\text{m}}_{\text{kin}}    + \CL_{\text{int}}  ~, 
\end{split}
\end{equation}
where $\CL^{\text{m}}_{\text{kin}}$ is the matter kinetic terms and $\CL_{\text{int}}$ involves an interactions of arbitrarily high derivative orders. At lowest order, the gauge field couples to matter fields through the conserved current, so that $\CL_{\text{int}}$ takes the general form
\begin{equation}
\begin{split}
\CL_{\text{int}} = - A^\mu J^M_\mu + \CL_{\text{int}}^{\text{HO}}~. 
\end{split}
\end{equation}
A simple argument \cite{Elvang:2016qvq,Dumitrescu:2015fej} implies that the leading soft photon theorem depends only on the lowest order interaction $ - A^\mu J^M_\mu$ and more generally, is related to the forward matrix element $\lim_{p'\to p}\bra{f,p,s} J^M_\mu(x) \ket{f',p',s'}$. The latter is fixed entirely by current conservation and Poincar\'e invariance (see for instance, Chapter 10 of \cite{Weinberg:1995mt}) and depends only on the $U(1)$ charge $Q$ of the state. For this reason, the leading soft factor is does not depend on the details of the theory and has the form
\begin{equation}
\begin{split}\label{leadingsofttheorem}
\lim_{\omega_s \to 0} \omega_s A_{n+1}(p_s,h_s) = \SS^\0 A_n~, \qquad \SS^\0 = e \omega_s \sum_{k=1}^n Q_k \frac{p_k \cdot \ve_s }{ p_k \cdot p_s } ~. 
\end{split}
\end{equation}
Here, $A_{n+1}$ is an $(n+1)$-point amplitude with soft photon momentum $p_s$ and helicity $h_s$ and $A_n$ is the amplitude without the soft particle. We use the now standard convention where all particles in an amplitude are taken to be outgoing. Incoming particles are then distinguished by negative $p^0$. 

In a similar way, the subleading soft factor depends on the next-to-forward matrix element $\lim_{p'\to p}\p_{p^\mu} \bra{f,p,s} J^M_\nu(x) \ket{f',p',s'}$. This is also fixed by current conservation and Lorentz invariance. We refer to this contribution to the subleading soft factor as the \emph{universal} piece. However, unlike the leading soft factor, the subleading soft factor also depends on higher derivative interactions $\CL_{\text{int}}^{\text{HO}}$ and we refer to this contribution as the \emph{non-universal piece}. The full subleading soft theorem takes the form,
\begin{equation}
\begin{split}\label{softth}
\lim_{\omega_s \to 0}   \big( 1 + \omega_s\p_{\omega_s}  \big)  A_{n+1} (p_s,h_s) = \SS^\1 A_n + {\tilde \SS}^\1  A_n ~. 
\end{split}
\end{equation}
$\SS^\1$ is the \emph{universal} piece which takes the form \cite{Low:1954kd,GellMann:1954kc,Low:1958sn,Burnett:1967km}, 
\begin{equation}
\begin{split}\label{universalfactor}
\SS^\1   = - i e \sum_{k=1}^n Q_k \frac{ p_s^\mu \ve^\nu_s }{ p_k \cdot p_s } \CJ_{k\mu\nu}~, 
\end{split}
\end{equation}
where $\CJ_{k\mu\nu}$ is the angular-momentum operator acting on $k$th particle. ${\tilde \SS}^\1$ is the \emph{non-universal} piece and takes the following general form \cite{Elvang:2016qvq},
\begin{equation}
\begin{split}\label{softfactor}
 {\tilde \SS}^\1_{+} A_n =   \sum_{k=1}^n \frac{ \savg{sk}   }{ \avg{s k} } \SF_k A_n ~, \qquad  {\tilde \SS}^\1_{-} A_n =  \sum_{k=1}^n \frac{ \avg{sk}   }{ \savg{s k} }\SF_k^\dagger A_n ~,
\end{split}
\end{equation}
Here, $\SF$ and $\SF^\dagger$ is a particle-changing operator. The subscript $k$ indicates that the operators act on the $k$th particle in $A_n$. The precise forms of $\SF$ and $\SF^\dagger$ depends on $\CL_{\text{int}}^{\text{HO}}$. 

For instance, the following higher derivative operator contributes to ${\tilde \SS}^\1$,
\begin{equation}
\begin{split}\label{scalarint1}
\CO_1 = \phi \big[ g_1 (F_{\mu\nu}^+)^2 + g_1^* ( F_{\mu\nu}^- )^2 \big] ~. 
\end{split}
\end{equation}
Here, $\phi$ is a real scalar (axion) and $F^\pm = \frac{1}{2} \big( F \mp i \ast F  \big)$. In this case,
\begin{equation}
\begin{split}\label{scalarcase1}
\bra{\phi,p} \SF   &=   2e^2 g_1 \bra{F,p,-}~,\quad \bra{F,p,+} \SF = - 2e^2 g_1\bra{\phi,p}~, \quad \bra{F,p,-} \SF = 0 ~, \\
\bra{\phi,p} \SF^\dagger &= 2e^2 g_1^* \bra{F,p,+} ~,~\, \bra{F,p,-} \SF^\dagger = - 2e^2 g_1^*  \bra{\phi,p} ~, ~\,\,\bra{F,p,+} \SF^\dagger = 0 ~.
\end{split}
\end{equation}
Here, $a_{f,s}(\vec{p}\,)$ is the annihilation operator that appears in the mode expansion $f(x)$ and creates an outgoing helicity $s$ state, $\bra{f,p,s} = \bra{0} a_{f,s}(\vec{p}\,)$. The creation and annihilation operators satisfy
\begin{equation}
\begin{split}\label{anorm}
\big[ a_{f,s}(\vec{p}\,) , a^\dagger_{f',s'}(\vec{p}\,') \big\} = (2\pi)^3 ( 2 \omega_p ) \delta_{f,f'} \delta_{s,s'} \delta^3 \big( \vec{p} - \vec{p}\,' \big) ~.
\end{split}
\end{equation}

Another example of a higher derivative operator that contributes to ${\tilde \SS}^\1$ is the magnetic dipole moment operator,
\begin{equation}
\begin{split}\label{spinorint1}
\CO_2 =  g_2 F^-_{\mu\nu} \psi_1 \sigma^{\mu\nu} \psi_2  + g^*_2 F^+_{\mu\nu} {\bar \psi}_1 {\bar \sigma}^{\mu\nu} {\bar \psi}_2~.  
\end{split}
\end{equation}
Here, $\psi_1$ and $\psi_2$ are two Weyl spinors. In this case
\begin{equation}
\begin{split}\label{spinorcase1}
\bra{{\bar \psi}_1,p,+} \SF &= \sqrt{2} i e g_2 \bra{\psi_2,p,-}~, \qquad\quad\,\bra{{\bar \psi}_2,p,+} \SF = -  \sqrt{2} i e g_2 \bra{\psi_1 ,p,-}~, \\
\bra{\psi_1,p,-} \SF &= 0 ~,\qquad\qquad\qquad\qquad\quad~~\,\, \bra{\psi_2,p,-} \SF = 0 ~, \\
\bra{\psi_1,p,-} \SF^\dagger &= - \sqrt{2} i e g^*_2 \bra{ {\bar \psi}_2,p,+}~, \qquad \bra{\psi_2,p,-} \SF^\dagger =  \sqrt{2} i e g^*_2  \bra{ {\bar \psi}_1 ,p,+}~, \\
\bra{{ {\bar \psi}}_1,p,+} \SF^\dagger &= 0 ~, \qquad\qquad\qquad\qquad\quad~\,\,  \bra{{ {\bar \psi}}_2,p,+} \SF^\dagger = 0 ~.
\end{split}
\end{equation}
Note that $\SF$ ($\SF^\dagger$) decreases (increases) the helicity of the state by 1. This is consistent with little group invariance of the subleading soft theorem.

We present a diagrammatic proof of \eqref{scalarcase1} and \eqref{spinorcase1} in Appendix \ref{app:softthproof}.

In this paper, we focus on the contributions of $\CO_1$ and $\CO_2$ to ${\tilde \SS}^\1$. However, there are other operators involving interactions of the gauge field with the gravitino and the graviton which also contribute to ${\tilde \SS}^\1$. These were classified in \cite{Elvang:2016qvq} and we list them here for completeness,
\begin{equation}
\begin{split}
\CO_3 &= g_3 \chi^\mu \sigma^\nu {\bar \psi} F_{\mu\nu} + c.c. ~,  \qquad \CO_4 = g_4 \chi^\mu \sigma^\nu {\bar \psi} {\tilde F}_{\mu\nu} + c.c ~, \qquad \CO_5 = g_5 h^{\mu\nu} T^{\text{em}}_{\mu\nu} ~. 
\end{split}
\end{equation}
where $T^{\text{em}}_{\mu\nu} = F_\mu{}^\a F_{\nu\a} - \frac{1}{4} g_{\mu\nu} F_{\a\b} F^{\a\b}$. 

As previously noted, a remarkable feature of ${\tilde \SS}^\1$ is its universal kinematic structure (which in \eqref{softfactor} appears through the square and angle brackets) despite the varied nature of the higher derivative interactions $\CO_i$ which give rise to it.

\section{Preliminaries}\label{sec:clreview}

\subsection{Notations and Conventions}

We consider a $U(1)$ gauge field $A$ coupled to massless matter fields $\varphi$ via ${\cal L}_{\text{int}}^{\text{HO}}$ and satisfying the field equations
\begin{equation}
\begin{split}\label{eom}
d \ast F  - d \ast G = e^2 \ast J^{M}  ~, 
\end{split}
\end{equation}
where $F = dA$. $J^{M}$ is the matter current associated to charged matter fields and $G_{\mu\nu}$ is a term that arises due to ${\cal L}_{\text{int}}^{\text{HO}}$. These equations are invariant under $U(1)$ gauge transformations,
\begin{equation}
\begin{split}\label{gaugetrans}
\delta_{ \lambda} A  = d { \lambda} ~, \qquad \delta_{ \lambda} \varphi = i Q_\varphi { \lambda} \varphi ~. 
\end{split}
\end{equation}
where $Q_\varphi$ is the $U(1)$ charge of the field $\varphi$.

In the absence of ${\cal L}_{\text{int}}^{\text{HO}}$ the charge that generates \eqref{gaugetrans} on a Cauchy surface $\Sigma$ is \cite{Strominger:2013lka}.
\begin{equation}
\begin{split}\label{electriccharge}
Q_\Sigma[\lambda] = \frac{1}{e^2} \int_{\p\Sigma} \lambda \ast F ~. 
\end{split}
\end{equation}
In addition to this, we will also be interested in the dual magnetic charge
\begin{equation}
\begin{split}\label{magneticcharge}
{\tilde Q}_\Sigma[\lambda] = \frac{1}{2\pi} \int_{\p\Sigma} \lambda F ~. 
\end{split}
\end{equation}
We normalize the charges so that they are integers for $\lambda = 1$.

In this paper, we will be interested in the charges on the null boundaries of Minkowski spacetime, $\ci^\pm$. These are best described in retarded coordinates
\begin{equation}
\begin{split}
ds^2 = - du^2 - 2 du dr + r^2 \g_{ab} d \Theta^a d\Theta^b ~, 
\end{split}
\end{equation}
where $\g_{ab}$ is the metric on the asymptotic $S^2$. In stereographic coordinates $\Theta^a = (z,\bz)$, $\g_{zz}=\g_{\bz\bz}=0$, $\g_{z\bz} = 2 (1+|z|^2)^{-2}$. 
These are related to the usual Cartesian coordinates by
\begin{equation}
\begin{split}
x^0 = u + r ~, \qquad x^1 + i x^2 = r \frac{2z}{1+|z|^2} ~, \qquad x^3 = r \frac{ 1 - |z|^2}{ 1 + |z|^2 } ~. 
\end{split}
\end{equation}
Future null infinity, $\ci^+$, is described by the limit $x^0 \to \infty$ keeping $(u,\Theta)$ fixed. Boundaries of $\ci^+$ are located at $u=\pm\infty$ keeping $\Theta$ fixed and are denoted by $\ci^+_\pm$. 

In the rest of this paper, we denote by $D_a$ the $\g$-covariant derivative. All Latin indices are raised and lowered w.r.t. $\g_{ab}$ and $D^2 = D^a D_a$. We denote by $\eta_{ab}$ the volume element on the asymptotic $S^2$. In stereographic coordinates $\eta_{z\bz} = i \g_{z\bz}$. 

\subsection{Large Gauge Transformations and Charges in QED}

The leading and subleading soft photon theorem in massless QED are associated with $\CO(1)$ and $\CO(r)$ large gauge transformations respectively \cite{Campiglia:2016hvg}. In this section, we review these gauge transformations and the corresponding charges.

To study the dynamics of the gauge field, we must fix a gauge. For our purposes, it is convenient to work in Lorenz gauge
\begin{equation}
\begin{split}
d \ast A  = 0 ~. 
\end{split}
\end{equation}
This gauge condition is preserved by gauge transformations generated functions $\lambda$ satisfying
\begin{equation}
\begin{split}\label{gaugecond}
d \ast d \lambda = 0 ~. 
\end{split}
\end{equation}

\paragraph{$\CO(1)$ gauge transformations} These are finite and non-zero on $\ci^+$ and have the asymptotic behaviour
\begin{equation}
\begin{split}\label{O1gaugecond}
\lambda(u,r,\Theta) = \ve(\Theta) + \CO(r^{-1})~. 
\end{split}
\end{equation}
The charges for these gauge transformation may then be constructed by substituting \eqref{O1gaugecond} into \eqref{electriccharge} and \eqref{magneticcharge} which gives
\begin{equation}
\begin{split}
Q_{\ci^+}[\ve]  &=  \frac{1}{e^2} \int_{\ci^+} du d^2 \Theta \sqrt{\g} \big[ D^a \ve F_{ua}^\0 - \ve J_u^{M\2} \big] ~, \\
{\tilde Q}_{\ci^+}[\tilde \ve] &= \frac{1}{2\pi} \int du d^2\Theta \sqrt{\g} \eta^{ab} \tilde \ve D_a F_{ub}^\0 ~. 
\end{split}
\end{equation}
Here, $f^{(n)}$ denotes the coefficient of $r^{-n}$ in the large $r$ expansion of the field $f(x)$. In order to determine these charges, we have repeatedly used the equations of motion \eqref{eom}. We refer the reader to \cite{Campiglia:2016hvg,He:2014cra} for details of this derivation. Similar charges may be constructed also on $\ci^-$.

\paragraph{$\CO(r)$ gauge transformations}  These diverge linearly in $r$ (or equivalently, $x^0$) near $\ci^+$ and have the asymptotic form
\begin{equation}
\begin{split}\label{divlargegauge}
\lambda(u,r,\Theta) = r \mu(\Theta) + \frac{u}{2} \big( D^2 + 2 \big) \mu(\Theta) + \CO(r^{-1})~. 
\end{split}
\end{equation}
The asymptotic charges  localized at ${\cal I}$ associated to such large gauge transformations cannot be intrinsically defined on ${\cal I}$ as they violate fall-off conditions on the radiative fields. However, as was shown in \cite{Campiglia:2016hvg} such charges can be defined via a (late time) limit of the charges defined on spatial slices through covariant phase space methods. The corresponding charges \eqref{electriccharge} and \eqref{magneticcharge} divergent piece that is proportional to the leading soft photon charge and hence vanishes due to the leading soft photon theorem. A few details of this procedure are presented in the next section. We finally have the charge
\begin{align}
Q^\1_{\ci^+}[\mu] &=  \frac{1}{2e^2} \int du d^2 \Theta \sqrt{\g}   \big[ u D^a D^2 \mu  F^\0_{ua} + e^2  D^a \mu  \big( u D_a J_u^{M\2}    -   J_a^{M\2} \big)  \big] ~, \label{elcharge} \\
{\tilde Q}^\1_{\ci^+}[\tilde \mu] &= \frac{1}{4\pi}  \int du d^2 \Theta \sqrt{\g}  \eta^{ab}  \big[   u D_a D^2 \tilde \mu F^\0_{ub} +  e^2 D_a\tilde  \mu  J_b^{M\2} \big]  ~. \label{magcharge}
\end{align}

For later use, it will be convenient to define charges $\CQ_{\ci^+}[\mu]$ and ${\bar \CQ}_{\ci^+}[\mu]$ where
\begin{equation}
\begin{split}\label{newchargedef}
\CQ_{\ci^+}[\mu] \equiv \frac{1}{2} Q^\1_{\ci^+}[\mu]  -  \frac{\pi i}{e^2} {\tilde Q}^\1_{\ci^+}[\mu]   ~. 
\end{split}
\end{equation}
In stereographic coordinates, the charges read
\begin{equation}
\begin{split}\label{charges1}
\CQ_{\ci^+}[\mu] &= \frac{1}{e^2} \int   d^2 z D_\bz^2  D^\bz \mu N_z^\1  + \frac{1}{2} \int du d^2 z D_z \mu \big[ u D_\bz J_u^{M\2} - J_\bz^{M\0} \big]  ~.  \\
\end{split}
\end{equation}
Here, we have define a photon zero-mode
\begin{equation}
\begin{split}\label{Nzonedef}
N_a^\1 &\equiv \int du   u  F_{ua}^\0  ~. 
\end{split}
\end{equation}
We separate the charges \eqref{charges1} into a soft piece, $\CQ^{S}_{\ci^+}[\mu]$ (linear in fields) and hard piece, $\CQ^{H}_{\ci^+}[\mu]$ (non-linear in fields). Note that the hard charges annihilate the perturbative vacuum state.

The charges \eqref{elcharge} and \eqref{magcharge} were shown in \cite{Campiglia:2016hvg} to be completely equivalent to the charge constructed in \cite{Lysov:2014csa} according to
\begin{equation}
\begin{split}\label{QYdef}
\frac{1}{2} Q^\1_{\ci^+}[\mu]  -  \frac{\pi i}{e^2} {\tilde Q}^\1_{\ci^+}[\tilde\mu] &= Q_{\ci^+}[Y] ~, \\
\end{split}
\end{equation}
where $Q_{\ci^+}[Y]$ is the charge constructed in
\cite{Lysov:2014csa}\footnote{In stereographic coordinates
\begin{equation}
\begin{split}
F_{ua}^\0 \big( \g^{ab} \g^{cd}  + \eta^{ab} \eta^{cd}  \big) D_b D_c Y_d = 2 F_{uz}^\0 D^z D_\bz Y^\bz + 2 F_{u\bz}^\0 D^\bz D_z Y^z ~. 
\end{split}
\end{equation}
},
\begin{equation}
\begin{split}
Q_{\ci^+}[Y] &= - \frac{1}{e^2} \int du d^2 \Theta \sqrt{\g} \left[ u F_{ua}^\0 \big( \g^{ab} \g^{cd}  + \eta^{ab} \eta^{cd}  \big) D_b D_c Y_d+ e^2 Y^a \big( u D_a J_u^\2  - J_a^\2 \big) \right]~.  
\end{split}
\end{equation}
and
\begin{equation}
\begin{split}\label{Yadef}
Y^a = - \frac{1}{4} \left( D^a \mu - i \eta^{ab} D_b {\tilde \mu} \right) ~. 
\end{split}
\end{equation}
Inversely, every two-dimensional vector field may be decomposed into a divergence and curl piece so that \eqref{Yadef} describes a general vector field on the asymptotic $S^2$. Thus, the two charges are indeed completely equivalent. Note that the charge $\CQ_{\ci^+}[\mu]$ in \eqref{newchargedef} corresponds to the choice $\mu = {\tilde \mu}$ (or equivalently to $Y^z = 0$) in \eqref{QYdef}. 

Analogous charges can also be constructed on $\ci^-$.

\subsection{Symplectic Structure for EFTs}

Having reviewed the large gauge charges in massless QED, we now discuss the corrections to the symplectic form and charges in the presence of higher derivative interactions, $\CL^{\text{HO}}_{\text{int}}$.

For simplicity, we start by considering massless scalar QED with charged scalar matter $\Phi$ alongwith a neutral axion $\phi$ which couples to the gauge field via\footnote{This corresponds to the operator $\CO_1$ \eqref{scalarint1} with $g_1 \in \mrr$.}
\begin{equation}
\begin{split}
L_{\textrm{int}}^{\text{HO}} = g_{1 }\phi F_{\mu\nu} F^{\mu\nu} ~. \label{simpint1}
\end{split}
\end{equation}

 The symplectic potential on the covariant phase space gets modified due to addition of the axion-photon interaction,
\begin{equation}
\begin{split}
\theta^{\mu}(\delta) &= \theta^\mu_{\textrm{old}}(\delta) + \theta^\mu_{\textrm{new}}(\delta) ~, \\
\theta^\mu_{\textrm{old}}(\delta) &= \sqrt{-g}\left[ \frac{1}{e^2} F^{\mu\nu}\delta A_{\nu} + ( \CD^{\mu} \Phi )^* \delta \Phi +  \delta \Phi^* ( \CD^\mu \Phi )  \right] ~, \\
\theta^\mu_{\textrm{new}}(\delta) &= - 4 \sqrt{-g}  g_1 \phi F^{\mu\nu} \delta A_\nu  ~. 
\end{split}
\end{equation}
Here, $\CD_\mu \Phi = \p_\mu \Phi - i Q A_\mu \Phi$ is the gauge covariant derivative. The full symplectic form is then
\begin{equation}
\Omega(\delta,\delta') =  \Omega_{\textrm{old}}(\delta,\delta') +  \Omega_{\textrm{new}}(\delta,\delta') ~. 
\end{equation}
If the variation of the fields $\delta A_{\mu}, \delta\phi, \delta\Phi$ satisfy the same fall-off conditions as the fields $A_\mu$, $\phi$ and $\Phi$ respectively, then it can be readily shown that on (future or past) null infinity
\begin{equation}
\Omega_{\cal I}(\delta,\delta^{\prime}) =  \Omega_{{\cal I}, \textrm{old}}(\delta,\delta^{\prime})~. 
\end{equation}
This is precisely the so-called \emph{radiative symplectic structure} on $\CI$.  It was shown in \cite{He:2014cra} that gauge transformations satisfying \eqref{O1gaugecond} preserve the fall-off conditions on the fields and hence in any EFT containing photons, the associated charges at ${\cal I}$ will only be determined by $\Omega_{\textrm{old}}$. As Ward identities associated to such charges were shown in \cite{He:2014cra} to be equivalent to Weinberg's soft photon theorem, we see that universality of this theorem is also a natural consequence of universality of radiative symplectic structure (and therefore the asymptotic charges) in any EFT containing photons.

\subsubsection{Subleading Electric Charge}

We now consider the case of subleading soft photon theorem in EFT which as we have discussed contains a universal as well as non-universal terms. The universal term (which is the only one present in tree level amplitudes of QED) it was shown in \cite{Lysov:2014csa, Campiglia:2016hvg} that this theorem is equivalent to Ward identities associated to a new symmetry of QED. These symmetries can be parameterized by vector fields on the conformal sphere or equivalently, by the class of divergent large gauge transformations \eqref{divlargegauge}.

We now follow \cite{Campiglia:2016hvg} consider the corrections to these charges in the presence of the interaction \eqref{simpint1}. After using the full symplectic structure and equations of motion, it can be shown (see Appendix \ref{electric-charge1}) that the charge is given by
\begin{equation}
\label{qlambda}
Q_{\Sigma_t}[\lambda] = \frac{1}{e^2}  \int_{\Sigma_t } dS_t \partial_{\mu}\left[ \sqrt{-g}\lambda (1 - 4 e^2 g_{1}\phi)F^{t\mu}\right] ~.
\end{equation}
We now want to evaluate these charges as $\Sigma_{t}$ approaches $\ci$, when the asymptotic behaviour of $\lambda$ is as in \eqref{divlargegauge}. For simplicity, we restrict ourselves to future null infinity which is obtained from $\Sigma_{t}$ by taking $t = u + r  \to \infty$ keeping $u,\Theta$ fixed. In the retarded co-ordinates $(u,r,\Theta)$ the integrand in \eqref{qlambda} is 
\begin{equation}
\label{rholambda}
\rho_{\lambda} = \sqrt{\g}\left[\partial_{r}(r^{2}\lambda\overline{F}_{ur}) - \partial_{u}(r^{2}\lambda\overline{F}_{ur})\right] + r^{2} \partial_a \big( \sqrt{\g}\lambda\overline{F}_u{}^a \big) ~.  
\end{equation}
where $\overline{F}_{\mu\nu} = \frac{1}{e^2} F_{\mu\nu} - 4 g_{1} \phi F_{\mu\nu}$. The last term in \eqref{rholambda} vanishes after integration over sphere. The standard boundary conditions for the axion field is $\phi  = \frac{\phi^{(1)}}{r} + \CO (r^{-1-\epsilon})$ with $\e>0$. Then, $\rho_{\lambda}$ is given by,
\begin{equation}
\label{rholambda}
\rho_\lambda = \sqrt{\g}\left[ \lambda^{(-1)}\overline{F}_{ur}^{(2)}\ - r\partial_{u} \big( \lambda^{(-1)}\overline{F}_{ur}^{(2)} \big) -\p_{u} \big( \lambda^{(-1)}\overline{F}_{ur}^{(3)} + \lambda^{(0)}\overline{F}_{ur}^{(2)} \big) \right] +\CO(r^{-\epsilon}) ~. 
\end{equation}

As shown in Appendix \ref{electric-charge1}, using equations of motion and Bianchi identities, the charge density is given by 
\begin{equation}\label{rholambdadiv}
\rho_\lambda = t \rho_{\text{div}}[\mu] +  \rho_{\text{finite}}[\mu] + \CO(t^{-\epsilon})   ~, 
\end{equation}
where $\rho_{\text{div}}[\mu]$ is a universal divergent term which is independent of the presence of irrelevant operators in QED and was derived in \cite{Campiglia:2016hvg},
\begin{equation}
\rho_{\text{div}}[\mu]=  \sqrt{\gamma}\mu\partial_{u}F_{ru}^{(2)}~. 
\end{equation}
The finite part of the charge density is given by 
\begin{equation}
\begin{split}
\CQ^\1_{\ci^+}[\mu] = \CQ^{\1,\text{old}}_{\ci^+}[\mu] + \CQ^{\1,\text{new}}_{\ci^+}[\mu] ~, 
\end{split}
\end{equation}
where
$\CQ^{\1,\text{old}}_{\ci^+}[\mu]$ is defined in \eqref{elcharge} and
\begin{equation}\label{elecchargenew}
\begin{split}
\CQ^{\1,\text{new}}_{\ci^+}[\mu] &= - 2g_{1}\int_{{\cal I}^{+}}du d^{2} \Theta \sqrt{\gamma}  D^a \mu\, \phi^{(1)} F_{ua}^\0 ~. 
\end{split}
\end{equation}

\subsubsection{Subleading Magnetic charge}

The asymptotic Magnetic charges associated to large gauge transformations were defined in \cite{Strominger:2015bla,Campiglia:2016hvg} and on a given Cauchy slice they are given by\footnote{
In analogy with eq.(\ref{qlambda}) it may be tempting to define a Magnetic charge where both terms in this equation are dualized. However, in the absence of Magnetic Monopoles, this will lead to an incorrect result as after dualizing both the terms in eq.(\ref{qlambda}) if we evaluate the resulting charge for $\lambda\ =\ 1$ it will not vanish (as it should in the case of zero Magnetic monopoles). 
We thus define the Magnetic charge in \emph{any EFT containing photons} as in eq.(\ref{qlambdatilde}).}
\begin{equation}\label{qlambdatilde}
\tilde{Q}_{\Sigma_t} [\tilde \lambda] = \int_{\Sigma_{t}}dS_{t}\partial_{\mu} \big(\tilde \lambda \eta^{t\mu\rho\sigma}{\cal F}_{\rho\sigma} \big)
\end{equation}

We can now compute this magnetic charge for $\CO(r)$ large gauge transformations. The corresponding charge density was computed in \S4 of \cite{Campiglia:2016hvg} and has the following asymptotic form.
\begin{equation}
\tilde \rho_{\tilde \lambda} = t \tilde\rho_{\text{div}}[\tilde \mu] +  \tilde\rho_{\text{finite}}[\tilde \mu] + \CO(t^{-\epsilon})   ~, 
\end{equation}
where
\begin{equation}
\begin{split}
\tilde\rho_{\text{div}}[\tilde \mu]   &= -\eta^{ab}\tilde{\mu}\partial_{u}{\cal F}^{(0)}_{ab} ~, \quad \tilde\rho_{\text{finite}}[\tilde \mu] = -\eta^{ab}\left[\tilde{\mu}\partial_{u} F^{(1)}_{ab} +  \frac{1}{2} D^2 \tilde{\mu} \big( u \partial_{u} + 1 \big) F^{(0)}_{ab} \right] ~. 
\end{split}
\end{equation}
As argued in \cite{Campiglia:2016hvg}, the divergent part of the asymptotic charge is the ``leading soft photon magnetic charge" and vanishes due to the magnetic soft photon theorem.

We can now analyze the finite part of the magnetic charge by using the Bianchi identity
\begin{equation}
F^{(1)}_{ab}  = 2 \p_{[a} F_{b] r}^{(2)}  ~,
\end{equation}
and the following equation of motion 
\begin{equation}
\partial_{u}F^{(2)}_{ar} = -\frac{1}{2} e^2  j_a^{M} - 2 e^2  g_{1} \phi^{(1)}F^{(0)}_{ua} - \frac{1}{2}\partial_a F_{ru}^{(2)} +\frac{1}{2}D^{b}F^{(0)}_{ab}~, 
\end{equation}
which implies
\begin{equation}
\partial_{u}F^{(1)}_{ab} = - e^2 \partial_{[a}j^{M\2}_{b]} - 4e^2  g_{1}\partial_{[a} \big( \phi^{(1)} F^{(0)}_{ub]} \big) - \frac{1}{2}D^2 F^{(0)}_{ab} ~. 
\end{equation}
Thus, the magnetic charge at $\CI^+$ is given by,
\begin{equation}
\begin{split}
\tilde \CQ^\1_{\ci^+}[\tilde \mu] = \tilde \CQ^{\1,\text{old}}_{\ci^+}[\tilde \mu] + \tilde \CQ^{\1,\text{new}}_{\ci^+}[\tilde \mu] ~, 
\end{split}
\end{equation}
where $\tilde \CQ^{\1,\text{old}}_{\ci^+}[\tilde \mu]$ is the universal part of the magnetic charge \eqref{magcharge} and $\tilde \CQ^{\1,\text{new}}_{\ci^+}[\tilde \mu] $ is the contribution to the magnetic charge from the axion-photon interaction, 
\begin{equation}\label{magchargenew}
\tilde \CQ^{\1,\text{new}}_{\ci^+}[\tilde \mu]  =  \frac{e^2 g_1}{\pi} \int du d^2 \Theta \sqrt{\g} \eta^{ab}\partial_{a}\tilde{\mu}\phi^{(1)} F^\0_{ub} .
\end{equation}

Finally, combining the \emph{new} electric and magnetic charges \eqref{elecchargenew} and \eqref{magchargenew} as in \eqref{newchargedef}, we find
\begin{equation}
\begin{split}
\CQ_{\ci^+}[\mu] = \CQ_{\ci^+}^{\text{old}}[\mu] + \CQ^{\text{new}}_{\ci^+}[\mu] ~,
\end{split}
\end{equation}
where $\CQ_{\ci^+}^{\text{old}}[\mu]$ is defined in \eqref{charges1} and
\begin{equation}
\begin{split}
\CQ^{\text{new}}_{\ci^+}[\mu] &=  - 2 g_1  \int du d^2 z D_z \mu  \phi^\1 F_{u\bz}^\0 ~. 
\end{split}
\end{equation}

Note also that the \emph{new} contributions to the charges may also be combined into a charge parameterized by a sphere vector field as in \eqref{QYdef} with the additional term being
\begin{equation}
\begin{split}
Q^{\text{new}}_{\ci^+} [Y] &= 4 g_1 \int du d^2 \Theta \sqrt{\g} \, Y^a \phi^\1 F_{ua}^\0 ~. 
\end{split}
\end{equation}

Similar magnetic charges may be analogously defined at past null infinity $\ci^-$.

\subsubsection{Generalization to Other Interactions}

In the previous two sections, we derive the subleading electric and magnetic charge in the special case of the higher derivative interaction $\CL_{\text{int}}^{\text{HO}} = g_1 \phi F_{\mu\nu} F^{\mu\nu}$. We now generalize, without details, our discussion and construct the charges for all the higher derivative interactions $\CO_i$.

The equations of motion in the presence of higher derivative interaction terms $\CO_i$ all take the form
\begin{equation}
\begin{split}
d \ast ( F - e^2 G ) = e^2 \ast J^M~. 
\end{split}
\end{equation}
where the precise form of $G$ depends on the interaction. 

For instance for the $\CO_1+\CO_2$ interaction
\begin{equation}
\begin{split}\label{Gmndef}
G_{\mu\nu} = 4 \phi \big( g_1 F_{\mu\nu}^+ + g_1^* F_{\mu\nu}^- \big) +  2 \big( g_2 \psi_1 {\sigma}_{\mu\nu} \psi_2 +  g^*_2 {\bar \psi}_1 {\bar \sigma}_{\mu\nu} {\bar \psi}_2\big) ~. 
\end{split}
\end{equation}
The corrections to the electric and magnetic charges \eqref{elcharge}, \eqref{magcharge} are now
\begin{equation}
\begin{split}
\CQ^{\1,\text{new}}_{\ci^+}[\mu] &= - \frac{1}{2} \int_{{\cal I}^{+}}du d^{2} \Theta \sqrt{\gamma}  D^a \mu G_{ua}^\1 ~, \\
\tilde \CQ^{\1,\text{new}}_{\ci^+}[\tilde \mu]  &= \frac{e^2  }{4\pi} \int du d^2 \Theta \sqrt{\g} \eta^{ab}\partial_{a}\tilde{\mu} G_{ub}^\1 ~, \\
\end{split}
\end{equation}
and finally
\begin{equation}
\begin{split}\label{chargefinal}
\CQ_{\ci^+}[\mu] &= \CQ^{\text{old}}_{\ci^+}[\mu] + \CQ^{\text{new}}_{\ci^+}[\mu] \\
&=  \frac{1}{e^2} \int   d^2 z D_\bz^2  D^\bz \mu N_z^\1  + \frac{1}{2} \int du d^2 z D_z \mu \big[ u D_\bz J_u^{M\2} - J_\bz^{M\0} - G_{u\bz}^\1  \big] ~. 
\end{split}
\end{equation}

\subsection{Mode Expansions}

The mode expansion for the gauge and matter fields are
\begin{equation}
\begin{split}
\phi(x) &=  \int \frac{d^3 q}{ ( 2\pi )^3 } \frac{1}{2\omega_q} \big[  a_\phi(\vec{q}\,) e^{i q \cdot x }  +   a^\dagger_\phi(\vec{q}\,) e^{-i q \cdot x }  \big]~, \\
\psi_{i\,\a} (x) &=  \int \frac{d^3 q}{ ( 2\pi )^3 } \frac{\sket{q}_\a }{2\omega_q}   \big[  a_{\psi_i,-}(\vec{q}\,) e^{i q \cdot x }  +   a^\dagger_{{\bar \psi}_i,+}(\vec{q}\,) e^{-i q \cdot x }  \big]~, \\
A_\mu (x) &= e \sum_{\a=\pm} \int \frac{d^3 q}{ ( 2\pi )^3 } \frac{1}{2\omega_q} \big[ \ve_\mu^{\a} (\vec{q}\,)^* a_{F,\a}(\vec{q}\,) e^{i q \cdot x }  + \ve_\mu^{\a} (\vec{q}\,) a^\dagger_{F,\a}(\vec{q}\,) e^{-i q \cdot x }  \big]  ~, \\
\end{split}
\end{equation}
The creation annihilation operators satisfy \eqref{anorm} and
\begin{equation}
\begin{split}
q \cdot \ve^{\a} (\vec{q}\,) = 0 ~, \qquad \ve_\mu^{+} (\vec{q}\,)^* = \ve_\mu^{-} (\vec{q}\,)  ~, \qquad \ve^{+} (\vec{q}\,) \cdot \ve^{-} (\vec{q}\,) = 1~, \qquad \ve^{\pm} (\vec{q}\,)^2 = 0~. 
\end{split}
\end{equation}

We now take a limit to $\ci^+$. For this purpose, it is convenient to parameterize the momenta as
\begin{equation}
\begin{split}\label{momentumparameter}
q^\mu = \omega_q ( 1 , {\hat x}_w ) ~, \qquad {\hat x}_w = \left( ~\frac{w+\bw}{1+w\bw} ~,  ~ \frac{-i(w-\bw) }{ 1 + w \bw } ~, ~  \frac{1 - w \bw }{ 1 + w \bw } ~ \right) ~. 
\end{split}
\end{equation}
With this choice, the polarization vectors are
\begin{equation}
\begin{split}
\ve^+_\mu(\vec{q}\,) = \frac{1}{\sqrt{2}} \left( - \bw , 1 , - i , - \bw \right) ~, \qquad \ve^-_\mu(\vec{q}\,) = \frac{1}{\sqrt{2}} \left( - w , 1 ,  i , - w \right) ~, 
\end{split}
\end{equation}
and
\begin{equation}
\begin{split}
 \bra{q}_\da = \sqrt{ \frac{ 2 \omega_q }{ 1 + w \bw } } \begin{pmatrix} - w \\ 1 \end{pmatrix}  = \sqrt{2\omega_q} \,  {\bar \xi}^+_\da(w,\bw)~. 
\end{split}
\end{equation}
Near $\ci^+$, the Weyl spinor takes the form \cite{Dumitrescu:2015fej}
\begin{equation}
\begin{split}
{\bar \psi}_{i\,\a} (u,r,\Theta) = \frac{1}{r} {\bar \psi}^\1_i(u,\Theta) {\bar \xi}^+_{\da} (\Theta) + \CO(r^{-2})~. 
\end{split}
\end{equation}

Then, near $\ci^+$, the stationary phase approximation localizes the angular integral over $w,\bw$ to $w\to z$. Explicitly,
\begin{equation}
\begin{split}\label{mainmodeexp}
\phi^\1 (u,\Theta) &=   - \frac{i}{ 8 \pi^2 }   \int_0^\infty d\omega_q    \left[ a_\phi    ( \omega_q {\hat x}_\Theta)  e^{ - i \omega_qu  }  - a_\phi^\dagger   (\omega_q {\hat x}_\Theta)  e^{ i \omega_qu  }  \right]    \,  , \\
\psi_i^\1 (u,\Theta) &=   - \frac{i}{ 8 \pi^2 }   \int_0^\infty d\omega_q \sqrt{2 \omega_q }    \left[ a_{\psi_i,-}    (\omega_q {\hat x}_\Theta)  e^{ - i \omega_qu  }  - a_{ {\bar \psi}_i ,+} ^\dagger   (\omega_q {\hat x}_\Theta)  e^{ i \omega_qu  }  \right] \, , \\
A_a^\0 (u,\Theta) &= -  \frac{i e}{8\pi^2}  E_a^\a  \int_0^\infty d\omega_q \left[ a_{F,\a}  (\omega_q {\hat x}_\Theta)    e^{- i \omega_q u }  - a_{F,\a}^\dagger (\omega_q {\hat x}_\Theta)  e^{  i \omega_q u    }   \right] \,  . \\
\end{split}
\end{equation}
Here, we have used the polarizations to define a zweibein on the asymptotic $S^2$ as
\begin{equation}
\begin{split}
E^\a_a  \equiv \lim_{r\to\infty} \frac{1}{r} \p_a x^\mu \ve_\mu^\a ~. 
\end{split}
\end{equation}
In stereographic coordinates,
\begin{equation}
\begin{split}
E_z^+ = E_\bz^- = \frac{ \sqrt{2} }{ 1 + z \bz } ~, \qquad E_z^- = E_\bz^+ = 0 ~. 
\end{split}
\end{equation}
Then, for the photon zero-mode \eqref{Nzonedef} is
\begin{equation}
\begin{split}
N_a^\1 (u,\Theta) =  \frac{ie}{8\pi} E_a^\a \lim_{\omega_s\to0} \p_{\omega_s}  \big[ \omega_s a_{F,\a} ( \omega_s {\hat x}_\Theta )  - \omega_s a_{F,\a}^\dagger ( \omega_s {\hat x}_\Theta )  \big]  ~. 
\end{split}
\end{equation}

In the parameterization \eqref{momentumparameter}, the commutator \eqref{anorm} reads
\begin{equation}
\begin{split}\label{anorm1}
\big[ a_{f,s}(\omega {\hat x}_\Theta ) , a^\dagger_{f',s'}(\omega' {\hat x}_{\Theta'} ) \big\} = \frac{2}{\omega} (2\pi)^3  \delta_{f,f'} \delta_{s,s'}  \delta ( \omega - \omega' ) \delta^2 ( \Theta , \Theta' )  ~.
\end{split}
\end{equation}
The $\delta$ function on the sphere is normalized to
\begin{equation}
\begin{split}
\int d^2 \Theta \sqrt{\g} \delta^2 ( \Theta , \Theta' )  = 1 ~. 
\end{split}
\end{equation}
In stereographic coordinates $ \delta^2 ( \Theta , \Theta' ) = \g^{z\bz} \delta^2 ( z - z' )$. 

The large $r$ expansions of the fields described here then imply
\begin{equation}
\begin{split}\label{Gmnlarger}
G_{u\bz}^\1 = 4 g_1 \phi^\1  F_{u\bz}^\0 + \sqrt{2} g_2 E_\bz^- \psi_1^\1 \psi_2^\1~. 
\end{split}
\end{equation}
where $G_{\mu\nu}$ is defined in \eqref{Gmndef}.

\section{Ward Identity $\to$ Soft Theorem}\label{sec:wardid}

We now derive the Ward identity corresponding to the subleading charges \eqref{charges1} and show that it is equivalent to the subleading soft theorem \eqref{softth}.

Invariance of the $\CS$-matrix under this symmetry\footnote{Alternatively, \eqref{wardidstart} is implied by a matching the field strength on $\ci^+$ with that on $\ci^-$ across spatial infinity $i^0$ \cite{Strominger:2017zoo}.} implies
\begin{equation}
\begin{split}\label{wardidstart}
\CQ_{\ci^+}[\mu] \CS - \CS \CQ_{\ci^-}[\mu]  = 0 ~. 
\end{split}
\end{equation}
To derive the Ward identity, we sandwich the above into $\bra{\text{out}} \cdots \ket{0}$ (the \emph{in} state is the vacuum since all particles are taken to be outgoing) and find
\begin{equation}
\begin{split}
2 \bra{\text{out}} \CQ^{S}_{\ci^+}[\mu] \CS \ket{0}    =  -  \bra{\text{out}}  \CQ^{H}_{\ci^+}[\mu] \CS \ket{0}  ~. \\ 
\end{split}
\end{equation}
Here, we use crossing symmetry and the fact that the hard charge annihilates the \emph{in} state. Using the explicit forms of the charges \eqref{chargefinal}, we find
\begin{equation}
\begin{split} 
&\int   d^2 w  E_w^+ D_\bw^2  D^\bw \mu  \lim_{\omega_s\to0}  \p_{\omega_s} \big[ \omega_s \bra{\text{out}}   a_{F,+} ( \omega_s {\hat x}_w )   \CS \ket{0} \big] = 4 \pi i e   \bra{\text{out}}  \CQ^{H}_{\ci^+}[\mu]  \CS \ket{0}  ~. \\
\end{split}
\end{equation}
where
\begin{equation}
\begin{split}\label{chargeofcharges}
& \CQ^{H}_{\ci^+}[\mu ] = \frac{1}{2} \int du d^2 w D_w \mu  \big[ u D_\bw J_u^{\2 M} - J_\bw^{\0 M} \big]  - \frac{1}{2} \int du d^2 w D_w \mu G_{u\bw}^\1~. 
\end{split}
\end{equation}
A similar soft theorem can be derived for the negative helicity photon.

We simplify further by setting $\mu = \mu_{(s)}$ where
\begin{equation}
\begin{split}
D_w \mu_{(s)} = E_\bw^- \frac{ \bz_s - \bw }{ z_s - w } \qquad \implies \qquad E_w^+ D_\bw^2 D^\bw \mu_{(s)} = 2\pi \delta^2 ( z_s - w )  ~. 
\end{split}
\end{equation}
Then,
\begin{equation}
\begin{split}\label{mainwardid}
&  \lim_{\omega_s\to0}  \p_{\omega_s} \big[ \omega_s   \bra{\text{out}}   a_{F,+} ( \omega_s {\hat x}_s )   \CS \ket{0} \big] = 2 i e   \bra{\text{out}}  \CQ^{H}_{\ci^+}[\mu_{(s)}]  \CS \ket{0}  ~.  \\
\end{split}
\end{equation}
We now analyze the RHS of \eqref{mainwardid}. The first term in \eqref{chargeofcharges} acts on one-particle states to reproduce to \emph{universal} part of subleading soft theorem \eqref{universalfactor} (see \cite{Lysov:2014csa,Campiglia:2016hvg,Conde:2016csj}) since
\begin{equation}
\begin{split}
& i e \bra{f , p , s} \int du d^2 w D_w \mu_{(s)} \big[ u D_\bw J_u^{\2 M} - J_\bw^{\0 M} \big] =   - i e  Q_k \frac{ p_s^\mu \ve^\nu_s }{ p \cdot p_s } \CJ_{\mu\nu} \bra{f , p , s} ~. 
\end{split}
\end{equation}
To complete our discussion, what remains is to show that the second term in \eqref{chargeofcharges} generates the \emph{non-universal} piece. To do this, we first use the mode expansions \eqref{mainmodeexp} which implies
\begin{equation}
\begin{split}
\int du G_{u\bw}^\1  &=  \frac{ E_\bw^-   }{2(2\pi)^3}   \int_0^\infty  d\omega \omega  \left(2  i e g_1  \big[ a_{F,+}^\dagger (\omega {\hat x}_w)   a_\phi    ( \omega {\hat x}_w )  - a_\phi^\dagger   (\omega {\hat x}_w)   a_{F,-}  (\omega {\hat x}_w)  \big] \right. \\
&\left. \qquad\qquad\qquad\qquad +   \sqrt{2} g_2 \big[ a_{\psi_1 , -}    (\omega {\hat x}_w)  a_{ {\bar \psi}_2 ,+} ^\dagger   (\omega {\hat x}_w)  + a_{ {\bar \psi}_1 ,+} ^\dagger   (\omega {\hat x}_w)   a_{\psi_2,-}    (\omega {\hat x}_w)  \big] \right).
\end{split}
\end{equation}
Then, using \eqref{anorm1}, we immediately find
\begin{equation}
\begin{split}
 -  i e \bra{f_k,p_k,s_k} \int du d^2 w D_w \mu_{(s)}  G_{u\bw}^\1 &= \frac{ \savg{sk} }{ \avg{sk} } \bra{f_k,p_k,s_k} \SF~. 
\end{split}
\end{equation}
Here, we have used the fact that under the parameterization \eqref{momentumparameter}, $\frac{ \savg{sk} }{ \avg{sk} } = - \frac{ \bz_s - \bz_k }{ z_s - z_k }$.

Thus, we see that the Ward identity \eqref{wardidstart} is implies the subleading soft theorem. 

\section{Soft Theorem $\to$ Ward Identity}\label{sec:invproof}

Above, we have shown that the Ward identity implies the soft theorem. To complete the equivalence, we need to show that the soft theorem implies the Ward identity. In this section, we do this for the special case of the subleading soft theorem where $A_n$ involves $n$ axions (coupled to the gauge field via \eqref{simpint1}). In this case, the soft theorem for a negative helicity soft photon reads
\begin{equation}\label{softth1}
\lim_{\omega_s \to 0}   \big( 1 + \omega_s\p_{\omega_s} \big) A_{n+1} (p_s,-) = {\tilde \SS}^\1_{-}  A_n ~. 
\end{equation}
We recall that 
\begin{equation}\label{softthaxion}
{\tilde \SS}^\1_{-}  A_n =  -\sum_{k=1}^{n} \frac{z_s  - z_k }{ \bz_s -  \bz_k} \SF_k^{\dagger} A_{n} ~. 
\end{equation}
Following the derivation in \cite{Lysov:2014csa} we apply $D_{z_{s}}^{2} {\hat \ve}_{\bz_s}^{+}$ on both sides of eq.(\ref{softthaxion}) and using the fact that 
\begin{equation}
\begin{split}
D_{z_{s}}^{2} \left( {\hat \ve}_{\bz_s}^{+}   \frac{z_s  - z_k }{ \bz_s -  \bz_k}  \right) &= \p_{z_{s}}^{2} \left( {\hat \ve}_{\bz_s}^{+} \frac{z_s  - z_k }{ \bz_s -  \bz_k}  \right)  - \Gamma^{z_s}_{z_sz_s}\p_{z_{s}} \left( {\hat \ve}_{\bz_s}^{+} \frac{z_s  - z_k }{ \bz_s -  \bz_k}  \right)   = \frac{ 2 \sqrt{2} \pi }{ 1 + |z|^2 } \delta^2 ( z - z_k ) ~. 
\end{split}
\end{equation}
we find
\begin{equation}
D_{z_{s}}^{2}  \big[ {\hat \ve}_{\bz_s}^{+}   \lim_{\omega_s \to 0}   (\ 1 + \omega_s\p_{\omega_s}) A_{n+1} (p_s,-) \big] = -   \frac{ 2 \sqrt{2} \pi }{ 1 + |z_s|^2 }  \sum_{k=1}^{n}  \delta^2 ( z - z_k )  \SF^{\dagger}_k A_n  ~. 
\end{equation}
Then multiplying by $Y^{z_s}$ and integrating over the sphere, we find
\begin{equation}
\int d^{2}z_{s}  D_{z_{s}}^{2}Y^{z_{s}}   {\hat \ve}_{\bz_s}^{+}  \lim_{\omega_s \to 0}   (\ 1 + \omega_s\p_{\omega_s}) A_{n+1} (p_s,-) = - 2\pi   \sum_{k=1}^{n}Y^{z_k} {\hat \ve}^-_{z_k} \SF^{\dagger}_k A_n ~. 
\end{equation}
Using the definition of $\SF^{\dagger}$ in \eqref{scalarcase1}, we can write the RHS of the above equation as  
\begin{equation}
\textrm{RHS}\ =\ \langle\textrm{out}\vert\ Q_{{\cal I}^{+}}[Y]S \ket{0} ~, \qquad Q_{{\cal I}^{+}}[Y] = 4g_{1} \int dud^{2}z Y^{z}\phi^{(1)} F^{(0)}_{uz} ~. 
\end{equation}

This along with the result of \cite{Lysov:2014csa} shows that the subleading soft theorem for negative helicity soft photons implies the Ward identity of asymptotic charges parametrized by sphere vector fields $Y = Y^{z}\partial_{z}$. Results of \cite{Campiglia:2016hvg} in conjunction with equivalence \eqref{QYdef} of charges then imply that the complete subleading soft theorem in tree-level amplitudes is equivalent to the Ward identity of $\CO(r)$ gauge transformations. 

Similar results can be obtained for positive helicity soft photons by considering charges parametrized by vector fields $Y = Y^{\bz}\partial_{\bz}$ as well as for more general amplitudes and interactions.

\section{Conclusions}

Following the seminal work in \cite{Strominger:2013jfa}, it is now becoming increasingly clear that universal soft theorems in gauge theories and gravity are statements regarding conservation of an infinite number of asymptotic charges in scattering processes. Sub-subleading theorem in quantum gravity and subleading theorems in gauge theories represent a puzzle in this regard as these theorems are not universal and their detailed structure depend on the on-shell three point function of the theory \cite{Laddha:2017ygw}. 

At the same time, the relationship between the subleading soft theorem in gauge theories and sub-subleading soft theorem in gravity is relatively less understood due to the fact that even at tree level and when only relevant operators are included in the action (\eg as in the case of QED), the generators of candidate asymptotic symmetries whose Ward identities are equivalent to the soft theorems diverge at $\ci$. These generators give rise to asymptotic charges which are naively divergent but are rendered finite due to the leading soft photon theorem. The resulting finite charges are equivalent to subleading soft photon theorem in QED.

In this paper we have shown that these statements remain true for the subleading soft photon theorem in effective field theories containing photons. More precisely, the divergent term in the charge vanishes universally due to the leading soft photon theorem and that the Ward identities of resulting finite charge is always equivalent to the full subleading soft photon theorem.

We believe that this result can be extended to sub-subleading theorem of tree-level quantum gravity amplitudes as well. Our results hint at a possible interconnection between non-universality of certain soft theorems and a large class of asymptotic symmetries which do not fit in the usual paradigm of symmetries or large gauge transformations due to divergent generators.

\section*{Acknowledgements}

PM is grateful to Thomas Dumitrescu, Daniel Kapec and Andrew Strominger for useful conversations. AL is indebted to Miguel Campiglia, Madhusudan Raman and Ashoke Sen for number of discussions on soft theorems in effective field theories. Work of PM is supported in part by DOE grant DE-SC0009988 and the Fundamental Laws Initiative at Harvard. Work of AL is supported in part by the Ramanujan Fellowship.

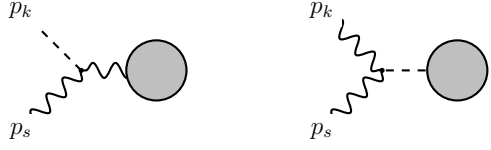
\appendix

\section{Diagrammatic Proof of Subleading Soft Photon Theorem}\label{app:softthproof}

In this section, we derive the higher derivative correction ${\tilde \SS}^\1$ to the subleading soft factor in the presence of $\CO_1$ and $\CO_2$. For this purpose, we may assume, for simplicity, that all fields are neutral.

These corrections are obtained from diagrams of the form
\begin{equation}
\begin{split}
\begin{tikzpicture}[thick,scale=0.8, every node/.style={scale=1}]
\node at (-3.5,-0.1) {$\lim\limits_{\omega_s \to 0} (1+\omega_s \p_{\omega_s} )~~~$};
\draw (0,0) -- (120:1.25);
\draw (0,0) -- (-120:1.25);
\draw (0,0) -- (-60:1.25);
\draw (0,0) -- (60:1.25);
\node at (180:1.6) {$p_s$};
\draw [decorate, decoration={snake, segment length=3mm, amplitude=1mm}] (0,0) -- (180:1.25);
\draw [line width=1pt,dotted] (-50:1) arc (-50:50:1);
\draw [fill=gray!50] (0,0) circle (0.5);
\node at (2,0) {$~\qquad=\,~\sum\limits_{k=1}^n$};
\begin{scope}[shift={(6,0)}]
\draw [dashed] (0,0) -- (180:1);
\draw [decorate, decoration={snake, segment length=3mm, amplitude=1mm}] ($(180:1)+(150:0.75)$) -- (180:1) ;
\draw (180:1) -- ($(180:1)+(-150:0.75)$);
\draw [fill=black] (180:1) circle (0.025);
\draw (0,0) -- (120:1.25);
\draw (0,0) -- (-120:1.25);
\draw (0,0) -- (-60:1.25);
\draw (0,0) -- (60:1.25);
\node at ($(180:1)+(-150:1)$) {$k$};
\node at ($(180:1)+(150:1)$) {$s$};
\draw [fill=gray!50] (0,0) circle (0.5);
\draw [line width=1pt,dotted] (-50:1) arc (-50:50:1);
\end{scope}
\end{tikzpicture}
\end{split}
\end{equation}

The interaction Lagrangian is
\begin{equation}
\begin{split}
\CL_{\text{int}} =  \text{Re}\,g_1  \phi  F_{\mu\nu} F^{\mu\nu}  + \text{Im}\,g_1  \phi F_{\mu\nu} (\ast F)^{\mu\nu}  + g_2 F_{\mu\nu} \psi_1 \sigma^{\mu\nu} \psi_2  + g^*_2 F_{\mu\nu} {\bar \psi}_1 {\bar \sigma}^{\mu\nu} {\bar \psi}_2~.  
\end{split}
\end{equation}
We start with the contribution of $\CO_1$ to ${\tilde \SS}^\1$. The $\phi\g\g$ vertex rule is
\begin{equation}
\begin{split}
\CV_{\phi\g\g} = 4 i \text{Re}\,g_1 \big[ p_{1\mu_2} p_{2\mu_1} - g_{\mu_1\mu_2} p_1 \cdot p_2   \big]  -  4i \text{Im}\,g_1 \ve_{\mu_1\mu_2\mu\nu} p_1^\mu p_2^\nu~. 
\end{split}
\end{equation}
Depending on the type of external particle to which the photon couples, there are two types of the diagrams
\begin{center}
\begin{tikzpicture}[thick,scale=0.8, every node/.style={scale=0.8}]
\draw  [decorate, decoration={snake, segment length=3mm, amplitude=1mm}]  (0,0) -- (-1,-1) node [fill=white] {$p_s$};
\draw  [decorate, decoration={snake, segment length=3mm, amplitude=1mm}]  (1.25,0) -- (0,0);
\draw [dashed] (0,0) -- (-1,1) node [fill=white] {$p_k$}; 
\draw [fill=black] (0,0) circle (0.025);
\draw [fill=gray!50] (1.25,0) circle (0.5);
\begin{scope}[shift={(5,0)}]
\draw  [decorate, decoration={snake, segment length=3mm, amplitude=1mm}]  (0,0) -- (-1,-1) node [fill=white] {$p_s$};
\draw [dashed] (1.25,0) -- (0,0);
\draw  [decorate, decoration={snake, segment length=3mm, amplitude=1mm}]   (0,0) -- (-1,1) node [fill=white] {$p_k$}; 
\draw [fill=black] (0,0) circle (0.025);
\draw [fill=gray!50] (1.25,0) circle (0.5);
\end{scope}
\end{tikzpicture}
\end{center}
These evaluate to
\begin{equation}
\begin{split}
A_{n+1} ( F, p_s , h_s \,; \phi,p_k ) ~ &\to ~  2 e^2  \big( \text{Re}g_1 + i h_s \text{Im} g_1   \big)  \sum_{h_k}  \frac{  M_s \cdot M_k }{p_s\cdot p_k}  A_n ( F , p_k , - h_k ) ~,  \\
A_{n+1} ( F, p_s , h_s \,; F , p_k , h_k ) ~ &\to ~ - 2 e^2 \big( \text{Re}\,g_1 + i h_s \text{Im}\, g_1 \big) \frac{ M_k \cdot M_s  }{ p_k \cdot p_s } A_n ( \phi , p_k ) ~. 
\end{split}
\end{equation}
where we have defined $M_{\mu\nu} \equiv p_{\mu} \ve_{\nu} - p_{\nu} \ve_{\mu}$ and $M_s \cdot M_k \equiv \frac{1}{2} M_s^{\mu\nu}M_{k\mu\nu} $. We note the property
\begin{equation}
\begin{split}\label{Mprops}
\ve_{\mu\nu\rho\sigma} M^{\rho\sigma} = - 2 i h M_{\mu\nu} ~. 
\end{split}
\end{equation}
Owing to \eqref{Mprops}, we note that $ M_s \cdot M_k$ is non-vanishing only if $h_k = h_s$,
\begin{equation}
\begin{split}
\frac{ M_s \cdot M_k}{ p_s \cdot p_k } =  \frac{\savg{sk} }{ \avg{sk} } ~~ \text{if}~h_s=h_k=1~, \quad \frac{ M_s \cdot M_k}{  p_s \cdot p_k }   =  \frac{\avg{sk} }{ \savg{sk} } ~~ \text{if}~h_s=h_k=-1~.
\end{split}
\end{equation}
For $h_s = +1$, we find
\begin{equation}
\begin{split}
A_{n+1} ( F, p_s , +\, ; \phi,p_k ) ~ &\to ~ 2 e^2 g_1  \frac{  \savg{sk} }{ \avg{sk} }  A_n ( F , p_k , - 1  )  ~, \\
A_{n+1} ( F, p_s , + \,; F , p_k , + ) ~ &\to ~ - 2 e^2  g_1 \frac{\savg{sk} }{ \avg{sk} }  A_n ( \phi , p_k ) ~, \\
A_{n+1} ( F, p_s , + \,; F , p_k , - ) ~ &\to ~  0 ~, \\
\end{split}
\end{equation}
and for $h_s = -1$
\begin{equation}
\begin{split}
A_{n+1} ( F, p_s , -\, ; \phi,p_k ) ~ &\to ~ 2 e^2 g^*_1  \frac{  \avg{sk} }{ \savg{sk} }  A_n ( F , p_k , + 1  )  ~, \\
A_{n+1} ( F, p_s , - \,; F , p_k , - ) ~ &\to ~ - 2 e^2  g^*_1 \frac{\avg{sk} }{ \savg{sk} }  A_n ( \phi , p_k ) ~, \\
A_{n+1} ( F, p_s , - \,; F , p_k , + ) ~ &\to ~ 0 ~. \\
\end{split}
\end{equation}
This reproduces the \eqref{softfactor} with \eqref{scalarcase1}.

Next, let us derive the contribution of $\CO_2$ to ${\tilde \SS}^\1$. The $\g\psi \psi$ vertices are
\begin{equation}
\begin{split}
\CV_{\g{\bar \psi}_1 {\bar \psi}_2} = 2 g^*_2 ( {\bar \sigma}^{\mu\nu} )_{\da\db} p_\nu  ~, \qquad \CV_{\g\psi_1\psi_2} = - 2 g_2 (\sigma^{\mu\nu})_{\a\b} p_\nu~.
\end{split}
\end{equation}
Again, depending on the type of external particle to which the photon couples, there are four types of diagrams
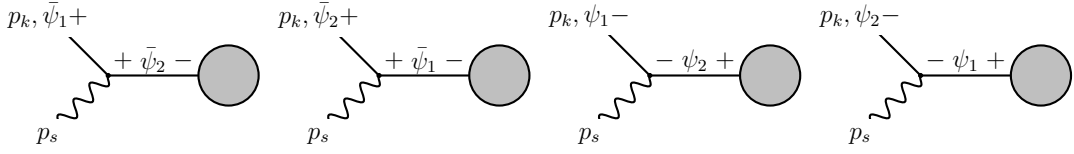
\begin{center}
\begin{tikzpicture}[thick,scale=0.8, every node/.style={scale=0.8}]
\begin{scope}[shift={(0,0)}]
\draw  [decorate, decoration={snake, segment length=3mm, amplitude=1mm}]  (0,0) -- (-1,-1) node [fill=white] {$p_s$};
\draw  (2,0) -- (0,0);
\node at ($(0.25,0)+(0,0.25)$) {$+$};
\node at ($(0.75,0)+(0,0.25)$) {${\bar \psi}_2$};
\node at ($(1.25,0)+(0,0.25)$) {$-$};
\draw  (0,0) -- (-1,1) node [fill=white] {$p_k,{\bar \psi}_1+$}; 
\draw [fill=black] (0,0) circle (0.025);
\draw [fill=gray!50] (2,0) circle (0.5);
\end{scope}
\begin{scope}[shift={(4.5,0)}]
\draw  [decorate, decoration={snake, segment length=3mm, amplitude=1mm}]  (0,0) -- (-1,-1) node [fill=white] {$p_s$};
\draw  (2,0) -- (0,0);
\node at ($(0.25,0)+(0,0.25)$) {$+$};
\node at ($(0.75,0)+(0,0.25)$) {${\bar \psi}_1$};
\node at ($(1.25,0)+(0,0.25)$) {$-$};
\draw  (0,0) -- (-1,1) node [fill=white] {$p_k,{\bar \psi}_2+$}; 
\draw [fill=black] (0,0) circle (0.025);
\draw [fill=gray!50] (2,0) circle (0.5);
\end{scope}
\begin{scope}[shift={(9,0)}]
\draw  [decorate, decoration={snake, segment length=3mm, amplitude=1mm}]  (0,0) -- (-1,-1) node [fill=white] {$p_s$};
\draw  (2,0) -- (0,0);
\node at ($(0.25,0)+(0,0.25)$) {$-$};
\node at ($(0.75,0)+(0,0.25)$) {$\psi_2$};
\node at ($(1.25,0)+(0,0.25)$) {$+$};
\draw  (0,0) -- (-1,1) node [fill=white] {$p_k,\psi_1-$}; 
\draw [fill=black] (0,0) circle (0.025);
\draw [fill=gray!50] (2,0) circle (0.5);
\end{scope}
\begin{scope}[shift={(13.5,0)}]
\draw  [decorate, decoration={snake, segment length=3mm, amplitude=1mm}]  (0,0) -- (-1,-1) node [fill=white] {$p_s$};
\draw  (2,0) -- (0,0);
\node at ($(0.25,0)+(0,0.25)$) {$-$};
\node at ($(0.75,0)+(0,0.25)$) {$\psi_1$};
\node at ($(1.25,0)+(0,0.25)$) {$+$};
\draw  (0,0) -- (-1,1) node [fill=white] {$p_k,\psi_2-$}; 
\draw [fill=black] (0,0) circle (0.025);
\draw [fill=gray!50] (2,0) circle (0.5);
\end{scope}
\end{tikzpicture}
\end{center}
These evaluate to
\begin{equation}
\begin{split}\label{A10}
A_{n+1} (F,p_s,h_s ; {\bar \psi}_1,p_k,+) ~&\to ~ - i e g_2 \frac{ \sbra{k} (\sigma \cdot M_s) \sket{k}    }{ p_k \cdot p_s }  A_n ( \psi_2 , p_k , - )  ~, \\
A_{n+1} (F,p_s,h_s ; {\bar \psi}_2,p_k,+) ~&\to ~ i e g_2 \frac{ \sbra{k} (\sigma \cdot M_s) \sket{k}    }{ p_k \cdot p_s }  A_n ( \psi_2 , p_k , - )  ~, \\
A_{n+1} ( F , p_s , h_s ; \psi_1 , p_k , -)  ~ & \to ~   - i e g_2^* \frac{  \bra{k} ( {\bar \sigma} \cdot M_s )   \ket{k} }{p_k \cdot p_s }  A_n ( {\bar \psi}_2 , p_k , + )~, \\
A_{n+1} ( F , p_s , h_s ; \psi_2 , p_k , -)  ~ & \to ~  i e g_2^* \frac{  \bra{k} ( {\bar \sigma} \cdot M_s )   \ket{k} }{p_k \cdot p_s }  A_n ( {\bar \psi}_1 , p_k , + ) ~. 
\end{split}
\end{equation}
Owing to \eqref{Mprops}, the first (last) two limits are non-vanishing only when $h_s=+1$ ($h_s=-1$) and
\begin{equation}
\begin{split}\label{A11}
\frac{ \sbra{k} (\sigma \cdot M_s) \sket{k}    }{ p_k \cdot p_s }   =  - \sqrt{2} \frac{ \savg{sk} }{ \avg{sk} } \quad \text{if}~ h_s = +1 ~, \\
 \frac{  \bra{k} ( {\bar \sigma} \cdot M_s )   \ket{k} }{p_k \cdot p_s }   =  \sqrt{2} \frac{ \avg{sk} }{ \savg{sk} } \quad \text{if}~ h_s = -1 ~. \\
\end{split}
\end{equation}
Substituting \eqref{A11} into \eqref{A10}, we reproduce \eqref{softfactor} with \eqref{spinorcase1}.

\section{Higher Derivative Contributions to the Symplectic Structure}\label{electric-charge1}

In this section, we derive the symplectic structure for the Lagrangian
\begin{equation}
\begin{split}
\CL = -\frac{1}{4e^2} \sqrt{-g} F_{\mu\nu} F^{\mu\nu} - \frac{1}{2}  \sqrt{-g}  ( \p_\mu \phi )^2  +  \sqrt{-g}  g_1 \phi F_{\mu\nu} F^{\mu\nu} ~. 
\end{split}
\end{equation}
Varying the Lagrangian, we find
\begin{equation}
\begin{split}\label{deltalag}
\delta \CL &=   \frac{1}{e^2}    \sqrt{-g}   \nabla^\mu \big( F_{\mu\nu} - 4 e^2 g_1 \phi F_{\mu\nu} \big)  \delta A^\nu   +  \sqrt{-g}  \left[ \nabla^2  \phi   + g_1 F_{\mu\nu} F^{\mu\nu}   \right]  \delta \phi   \\
&\qquad \qquad \qquad \qquad \qquad - \p_\mu \left[  \sqrt{-g}  \frac{1}{e^2} \big( F^{\mu\nu} - 4 e^2 g_1 \phi F^{\mu\nu} \big)   \delta A_\nu +  \sqrt{-g}  \p^\mu \phi \delta \phi \right]  ~. 
\end{split}
\end{equation}
The equations of motion for the gauge field are
\begin{equation}
\begin{split}\label{eom}
 \nabla^\mu {\overline F}_{\mu\nu}  = 0  ~, \qquad {\overline F}_{\mu\nu} = \frac{1}{e^2} F_{\mu\nu} - 4 g_1 \phi F_{\mu\nu} ~. 
\end{split}
\end{equation}
In addition to this, we have the Bianchi identities, $dF = 0$. At large $r$, these equations take the form
\begin{equation}
\begin{split}\label{eom1}
\p_u  F_{ur}^\2 &= D^a  F_{ua}^\0 ~, \\
\p_u F_{ur}^\3   &= D^a F_{ua}^\1 + 4 e^2 g_1 \p_u \big( \phi^\1 F_{ur}^\2 \big)  - 4 e^2 g_1 D^a \big( \phi^\1 F_{ua}^\0 \big)  ~, \\
\p_u   F_{ra}^\2 &=   F_{ua}^\1  + D^b F_{ba}^\0 - 4 e^2 g_1 \phi^\1 F_{ua}^\0 ~, \\
\p_u F_{ra}^\2 &= - F_{ua}^\1 - D_a F_{ur}^\2 ~. \\
\end{split}
\end{equation}
These equations imply
\begin{equation}
\begin{split}\label{maineq1123}
\p_u D^a F_{ua}^\1 =  - \frac{1}{2}  D^2 D^a  F_{ua}^\0   + 2 e^2 g_1 \p_u D^a  \big( \phi^\1 F_{ua}^\0 \big) ~. 
\end{split}
\end{equation}

From the boundary term of \eqref{deltalag}, we read off the symplectic potential
\begin{equation}
\begin{split}
\theta^\mu(\delta) &=  \sqrt{-g} \left[  \, {\overline F}^{\mu\nu}  \delta A_\nu +  \p^\mu \phi \delta \phi \right]  ~. \\
\end{split}
\end{equation}
The symplectic form is then
\begin{equation}
\begin{split}
\Omega_\Sigma (\delta , \delta' )  = \int_\Sigma d\Sigma_\mu \sqrt{-g}  \left[  \delta {\overline F}^{\mu\nu}  \delta' A_\nu -  \delta'  {\overline F}^{\mu\nu} \delta A_\nu \right]  ~. 
\end{split}
\end{equation}
The correction to the charge that generates large gauge transformations is given by $\delta Q_\Sigma[\lambda] =  \Omega_\Sigma (\delta , \delta_\lambda )  $. This is integrable and we find
\begin{equation}
\begin{split}
Q_\Sigma[\lambda] =  \int_\Sigma d\Sigma_\mu \sqrt{-g}  \, {\overline F}^{\mu\nu} \p_\nu \lambda =  \int_\Sigma d\Sigma_\mu \p_\nu \big[\sqrt{-g} \lambda \, {\overline F}^{\mu\nu}  \big]  ~. 
\end{split}
\end{equation}
In the last equality, we have integrated by parts and used the equations of motion \eqref{eom}. The charge \eqref{qlambda} is then obtained by setting $\Sigma_\mu = \delta_\mu^t$.

We now explicitly determine this charge for the $\CO(r)$ gauge transformations \eqref{divlargegauge} and reproduce \eqref{rholambdadiv}. We recall the integrand \eqref{rholambda},
\begin{equation}
\rho_\lambda = \sqrt{\g}\left[ \lambda^{(-1)}\overline{F}_{ur}^{(2)}\ - r\partial_{u} \big( \lambda^{(-1)}\overline{F}_{ur}^{(2)} \big) -\p_{u} \big( \lambda^{(-1)}\overline{F}_{ur}^{(3)} + \lambda^{(0)}\overline{F}_{ur}^{(2)} \big) \right] +\CO(r^{-\epsilon}) ~. 
\end{equation}
At large $r$
\begin{equation}
\begin{split}
{\overline F}_{ur}^\2 = F_{ur}^\2 ~, \qquad {\overline F}_{ur}^\3 = F_{ur}^\3 - 4 e^2 g_1 \phi^\1 F_{ur}^\2 ~. 
\end{split}
\end{equation}
Then, using \eqref{eom1} and \eqref{divlargegauge}, we find that up to terms that vanish upon integration
\begin{equation}
\begin{split}
\rho_\lambda &=   t \frac{1}{e^2}  \sqrt{\g}  \mu \p_u   {F}_{ru}^{(2)}  + \sqrt{\g} \left[  - \frac{1}{2e^2}  \mu  u D^2 D^a  F_{ua}^\0 + 2 g_1 \mu D^a \big( \phi^\1 F_{ua}^\0 \big) \right] +\CO(r^{-\epsilon}) ~.
\end{split}
\end{equation}
The new contribution to the charge is then
\begin{equation}
\begin{split}
\CQ^{\1,\text{new}}_{\ci^+}[\mu] = - 2 g_1  \int du d^2 \Theta \sqrt{\g} D^a   \mu  \phi^\1 F_{ua}^\0 ~. 
\end{split}
\end{equation}
which reproduces \eqref{elecchargenew}.

\bibliography{LM-bib}
\bibliographystyle{utphys}

\end{document}